\documentclass[aps,prl,twocolumn,superscriptaddress]{revtex4}

\usepackage{graphicx}
\usepackage{amsmath}

\advance \textheight by 5\bigskipamount

\begin{document}

\title{Structural tunability in metamaterials}

\author{Mikhail Lapine}
\email[Corresponding author:]{mlapine@uos.de}
\affiliation{Nonlinear Physics Center, Research School of Physics and Engineering, Australian National University, Australia}
\affiliation{Dpto.\ Electronica y Electromagnetismo, Facultad de Fisica, Universidad de Sevilla, Spain}

\author{David Powell}
\affiliation{Nonlinear Physics Center, Research School of Physics and Engineering, Australian National University, Australia}

\author{Maxim Gorkunov}
\affiliation{Institute of Crystallography, Russian Academy of Sciences, Moscow, Russia}

\author{Ilya Shadrivov}
\affiliation{Nonlinear Physics Center, Research School of Physics and Engineering, Australian National University, Australia}

\author{Ricardo Marqu\'es}
\affiliation{Dpto.\ Electronica y Electromagnetismo, Facultad de Fisica, Universidad de Sevilla, Spain}

\author{Yuri Kivshar}
\affiliation{Nonlinear Physics Center, Research School of Physics and Engineering, Australian National University, Australia}


\begin{abstract}
We propose a novel approach for efficient tuning of the transmission characteristics
of metamaterials through a continuous adjustment of the lattice structure, and confirm
it experimentally in the microwave range. The concept is rather general and
applicable to various metamaterials as long as the effective medium description
is valid.
The demonstrated continuous tuning of metamaterial response is highly desirable
for a number of emerging applications of metamaterials
including sensors, filters, switches, realizable in a wide frequency range.
\end{abstract}

\pacs{}

\maketitle

Metamaterials are prominent for the exceptional opportunities they offer in tailoring
macroscopic properties through appropriate choice and arrangement of
their structural elements~\cite{ML7,Ari7}. In this way, it is not only possible to design
a metamaterial for a required functionality, but also to implement further adjustment
capabilities at the level of assembly. This makes metamaterials different from conventional
materials and opens exciting opportunities
of multi-functionality via tunability.

Tunable metamaterials imply the ability to continuously change their properties
through an external influence or signal with the intrinsic mechanism of tunability.
The key means of tuning resonant metamaterials, naturally, lies in affecting the
system so as to change the parameters of the resonance. As a consequence, the characteristics of metamaterial
can be varied, enabling, for instance, tunable transmission.

The first approach to realize tunable metamaterials  based on nonlinear properties~\cite{Gor4}
has already been proven experimentally~\cite{PowShaKiv7,ShaKozWei8}, and further methods have been
suggested, e.g. based on reconfigurability of liquid crystals \cite{GorOsi08}.
However, such methods become increasingly difficult to implement at higher frequencies.

In this Letter, we put forward an approach which relies on the structural tuning of the entire metamaterial,
and is, conceptually, independent of the specific realization as well as scalable to any frequency provided that
macroscopic requirements for metamaterials are observed.

We explain the general principle of the proposed tuning method through a simple analogy.
Indeed, the properties of crystals are known to be determined by the nature of constituent
atoms as well as by the geometry of the crystal lattice, so in natural materials the collective
response of atoms determines the overall response to external fields~\cite{LL}.
In natural materials however, possibilities to tune their properties dynamically are limited
to naturally available crystals and yield relatively weak effects, such as electro-/magnetostriction,
photorefraction, etc. 

In contrast, metamaterials offer a unique opportunity to design and vary the structure
enabling a desired response function and a convenient mechanism for tunability.
More importantly, the range of tunability for a given property can be much broader
than in natural materials, as the lattice effects can be made much stronger
through higher efficiency of collective effects in the lattice, achieved
by an appropriate design.

 \begin{figure}[b]
 \includegraphics[width=\columnwidth]{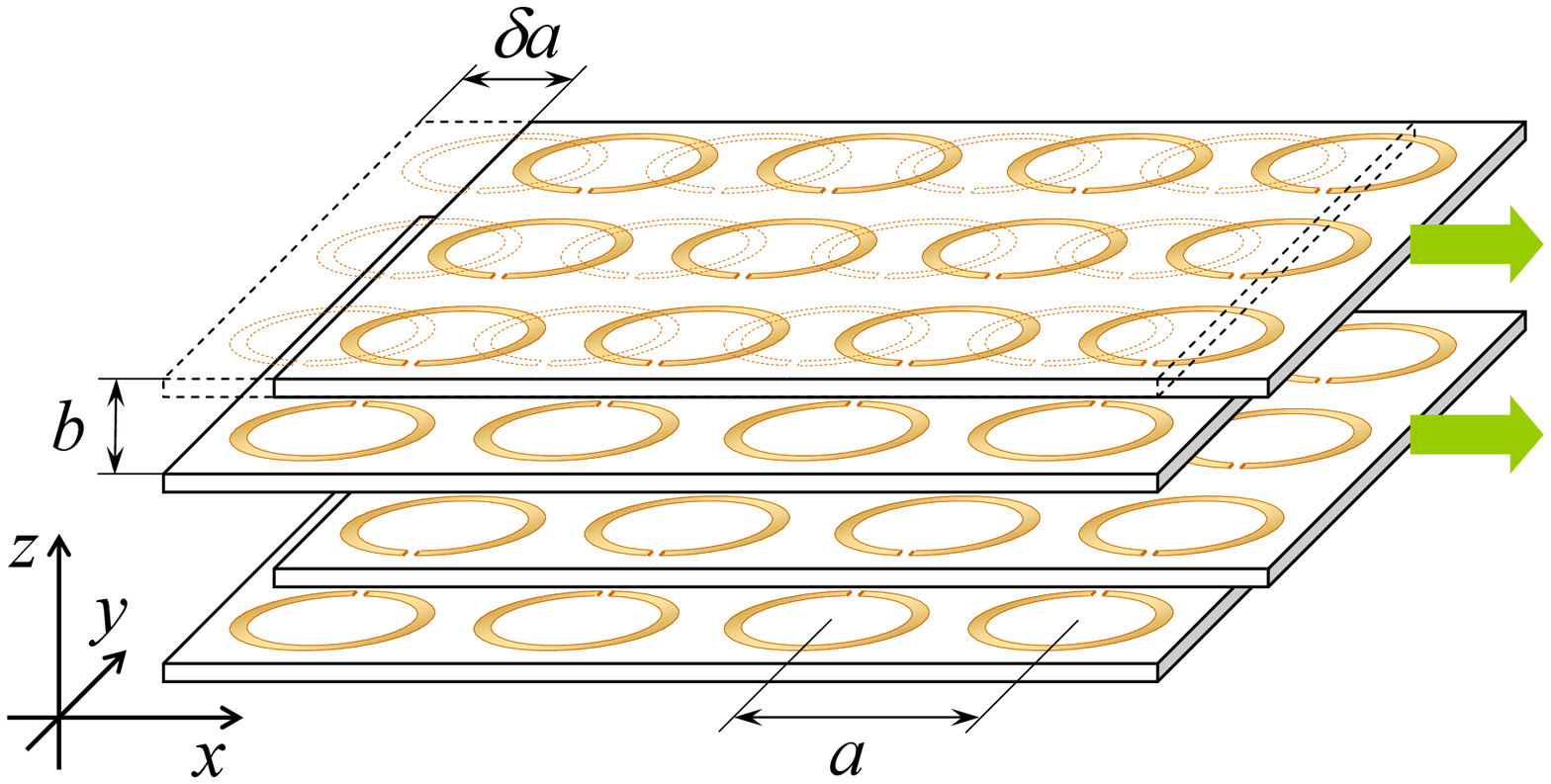}%
 \caption{
 Schematic of the staggered lattice shift with a lateral displacement of every second metamaterial layer.
 \label{sxem}}
 \end{figure}

To demonstrate the efficiency of this approach, we consider an anisotropic metamaterial
based on resonant elements suitable for providing artificial magnetism,
such as split-ring resonators of various kind, as shown in Fig.~\ref{sxem}.
For sufficiently dense arrays, the interaction between such elements differs considerably
from a dipole approximation, and the specific procedure to calculate the effective permeability
was developed earlier~\cite{glsr02}; the latter converges correctly to a Clausius-Mossotti
approximation in the limit of a sparse lattice. Consequently, the effect of mutual coupling is
enhanced dramatically as compared to conventional materials, and therefore
it is particularly suitable to demonstrate the efficiency of lattice tuning.

Accordingly, if all the characteristic dimensions (lattice constants and element size)
are much smaller than the wavelength, we can describe a regular lattice of such elements
by the resonant effective permeability
\begin{equation}
 \mu(\omega) = 1 - \frac{A \omega^2}{\omega^2 - \omega_r^2 + i \Gamma \omega }
 \label{muom}
\end{equation}
with the resonant frequency
\begin{equation}
 \omega_r = \omega_o \left( \frac{L_\Sigma }{L} + \frac{\mu_o \nu S^2}{3 L} \right)^{-1/2}
 \label{refreq}
\end{equation}
determined by both the properties of individual elements, such as the resonance frequency of a single element $\omega_o$, their geometry (which defines self-inductance $L$ and effective cross-section $S$), and concentration $\nu$, as well as their arrangement. The latter effect is determined by mutual interaction between the elements,
which in most practically realizable cases is defined by
\begin{equation}
 L_\Sigma = L + \mu_o r \Sigma,
 \label{lsig}
\end{equation}
where the so-called lattice sum $\Sigma$ can be calculated for a
given geometry of elements and their arrangement
through mutual inductance
\begin{equation}
 \sum_{n' \neq n } L_{nn'} (\omega) = - i \omega \mu_o r \cdot \Sigma,
 \label{sumsig}
\end{equation}
between all the elements in a physically small volume where the average
macroscopic field is evaluated \cite{glsr02}.

Note that the collective behavior of a number of elements in the lattice plays a crucial role, so that in dense arrays
the mutual interaction cannot be reduced to the approximation of nearest
neighbors adopted in recent rigorous models which account for spatial dispersion
\cite{BJMS08}.

 \begin{figure}
 \includegraphics[height=0.47\columnwidth,clip]{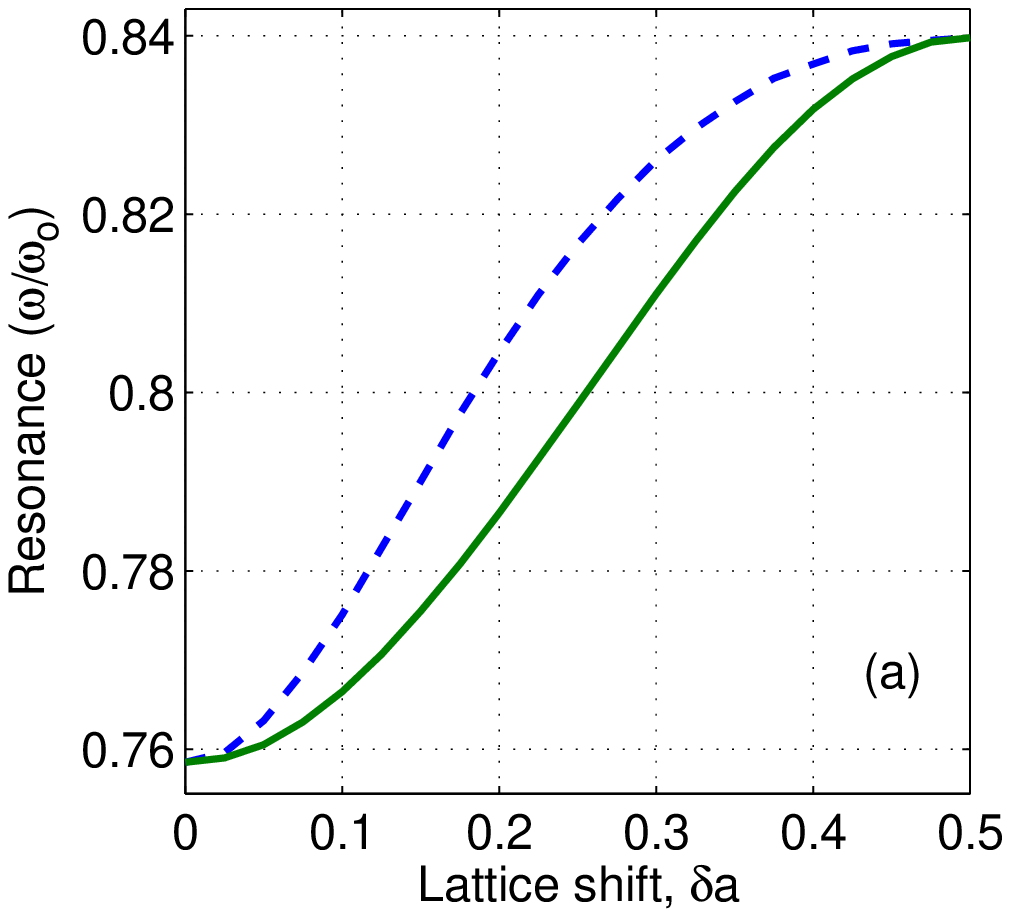}\hspace{0mm}%
 \includegraphics[height=0.47\columnwidth,clip]{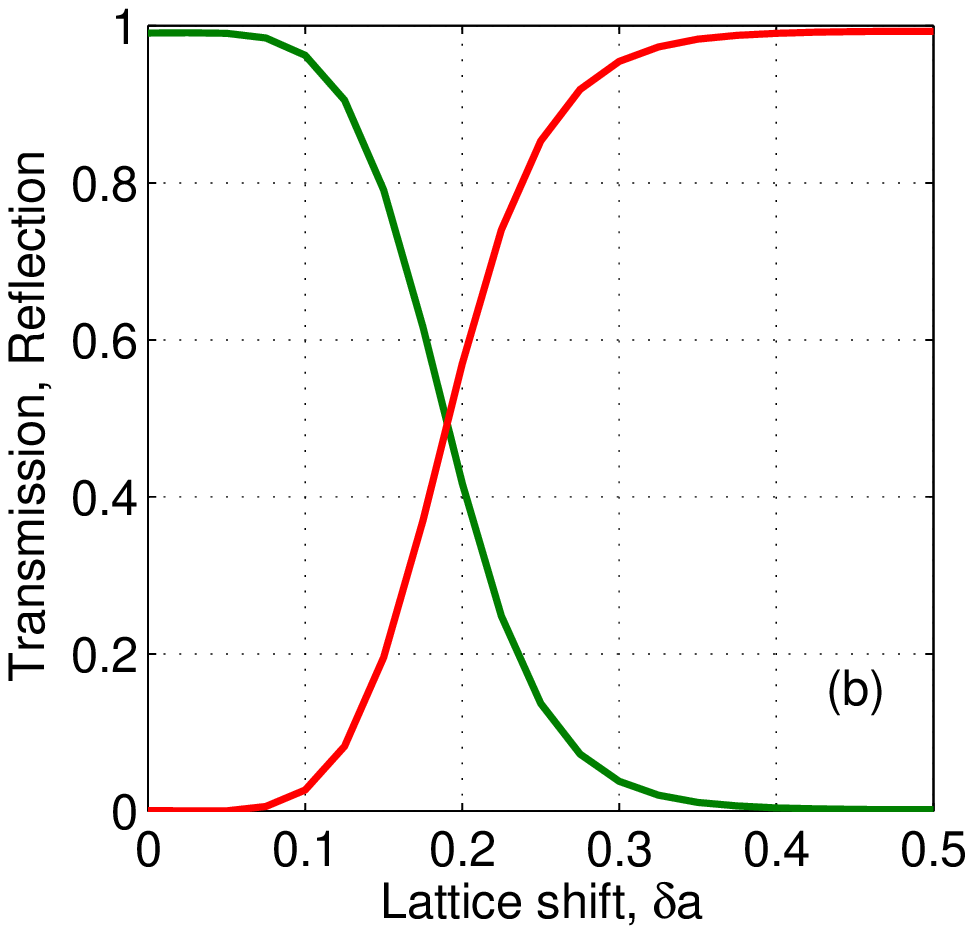}
 \caption{(a) Theoretical shift of the resonance frequency for continuous (dashed)
and staggered (solid) lattice shift strategy; (b) Calculated transmission and reflection
through a metamaterial slab (one wavelength thick) depending on lattice shift (staggered) at $\omega = 0.96\omega_o$.\label{teorsh}.}
 \end{figure}

The most straightforward lattice tuning approach is to vary the lattice constant $b$. We have shown \cite{glsr02}
that the resonance frequency can be remarkably shifted this way, and confirmed this with microwave experiments \cite{ShaPowMor7}. Accordingly, a slab of metamaterial can be tuned between transmission,
absorption and reflection back to transmission.
A clear disadvantage of this method is that varying $b$ implies a corresponding significant change
in the overall dimension of the metamaterial along $z$, which is undesirable for applications.

Here we propose another method of structural tuning, by means of a periodic lateral displacement of
layers in the $xy$ plane, so that the resonators become shifted along $x$ ($y$, or both)
by a fraction of the lattice constant $\delta a$ per each $b$ distance from a reference layer
with respect to the original position.
This decreases the overall mutual inductance in the system (Eq.\,\eqref{lsig})
and leads to a gradual increase in resonant frequency,
with a maximal effect archived for displacement by 0.5$a$ [see Fig.~\ref{teorsh}(a)].
Clearly, further shift is equivalent to smaller shift values until
the lattice exactly reproduces itself for the shift by $a$.
As a consequence, the resonance of the medium can be ``moved'' across a signal frequency,
leading to a drastic change in transmission characteristics [see Fig.~\ref{teorsh}(b)].
It is clear that for practical applications it is not even necessary to exploit
the whole range of lateral shift --- in the above example it is sufficient
to operate between $0.1a$ and $0.3a$ where most of the transition occurs.

 \begin{figure}[b]
 \includegraphics[width=\columnwidth]{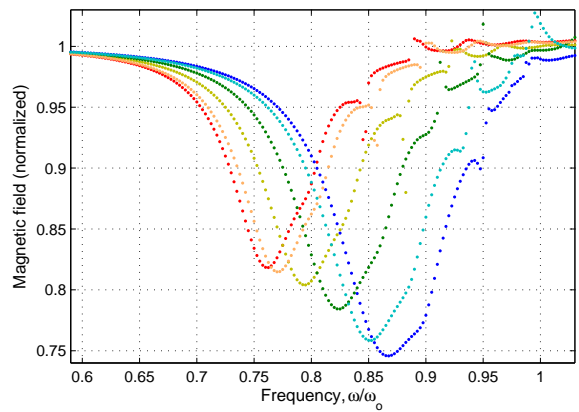}%
 \caption{Numerically calculated magnetic field beneath a finite metamaterial slab ($5 \times 1 \times 30$ elements)
 for a plane-wave incidence.
 Curves with dips from left to right correspond to increasing lattice shift from 0 to 0.5$a$.\label{finmod}}
 \end{figure}

Within the effective medium paradigm, a continuous shift of each layer
with respect to the previous one appears to provide maximal efficiency.
For finite samples, however, this poses certain disadvantages.
Indeed, for small $b/a$ ratio (which produces a stronger effect) even a small $\delta a$
shift would imply a remarkable inclination of the sample interface.
This would lead to undesirable shape distortion and cause
excitation of additional standing waves.
Preliminary experiments with this kind of tuning have shown
that the system generally features the predicted behavior, however
is rather unrepeatable with regards to excitation and measurement methods,
so the performance cannot be reliably assessed.

To overcome this difficulty, we consider a staggered lateral shift as shown in Fig.~\ref{sxem},
when every second layer is shifted while the rest of the structure remains at the original
position. This configuration leads to a slightly different efficiency pattern
[compare the two curves in Fig.~\ref{teorsh}(a)], but is equally useful; obviously, the two tuning
strategies converge to the identical result for $0.5a$ shift, as the lattice patterns shifted
with either method coincide in this case. For finite samples, the staggered shift is advantageous
as it keeps the sample interface straight and parallel to the axis of resonators at
all times, while slight regular distortion of the interface shape is not
expected to deteriorate the performance, provided that the overall number
of elements is sufficiently large so that surface effects are negligible.

For the experimental verification, we opted for a small reconfigurable system,
built up of single-split rings (2.25\,mm mean radius, 0.5\,mm strip width, 1\,mm gap)
printed with period $a = 7$\,mm on 1.5\,mm thick circuit boards. We have 5 resonators
in the propagation direction $x$ and only one period along $y$; 30 boards are stacked together
in $z$ direction with the minimum possible lattice constant $b = 1.5$\,mm used for the measurements.
The estimated resonance frequency of a single resonator 
is about 4.9\,GHz, however the resonance of the dense metamaterial is significantly shifted
to lower frequencies. To minimize the undesirable bianisotropic effects occuring in single-split rings,
the boards are assembled so that the gaps are oppositely oriented in adjacent layers (Fig.~\ref{sxem}),
resembling the design of broadside-coupled split-ring resonators~\cite{Ricardo}.
Transmission measurements (Rohde and Schwarz ZVB network analyzer)
were performed for various lattice shifts in WR-229 rectangular waveguide.

 \begin{figure}
 \includegraphics[width=\columnwidth]{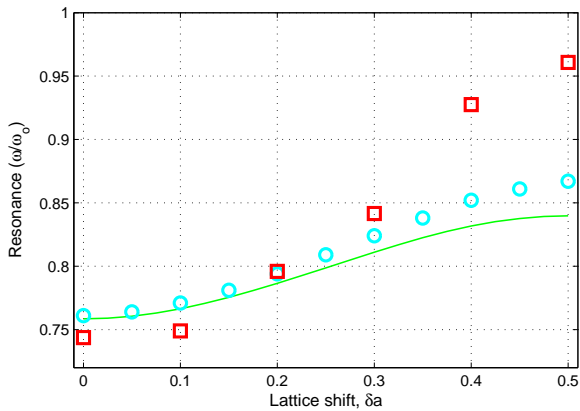}%
 \caption{Comparison between theoretical results and experimental data for
the resonance frequency shift:
Effective medium approach (solid); finite model (circles);
experimental results (squares).
 \label{compar} \label{rfsh}}
 \end{figure}

Note that the above system of $5 \times 1 \times 30$ resonators cannot be
described by an effective medium approach, as the number of elements is small.
Also, the system is not sufficiently subwavelength for a quasi-static approach to be used and spatial dispersion
becomes remarkable~\cite{Sim,Agran}. For this reason, we also perform semi-analytical calculations for the corresponding finite structures, with all the mutual inductances included (Eq.\,\eqref{sumsig}),
taking retardation effects into account.
Although the particular resonance values obtained this way (Fig.~\ref{finmod}), are different
from those which would be observed in a medium, the overall effect of lattice
tuning was qualitatively the same and predicts excellent performance (Fig.~\ref{compar}).

\begin{figure}[t]
 \includegraphics[width=\columnwidth]{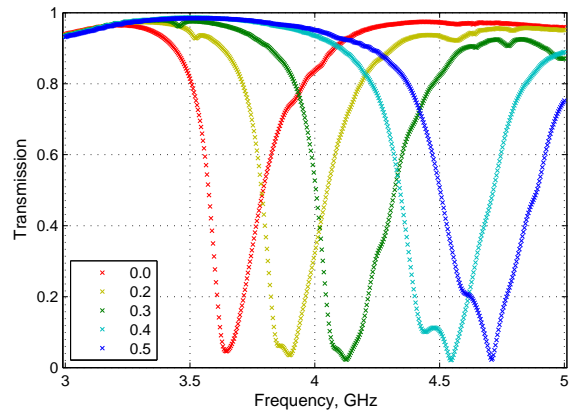}%
 \caption{Experimental transmission in a waveguide with metamaterial slab at different shifts.
 Curves with dips from left to right correspond to increasing lattice shift.
 \label{expbsc}}
 \end{figure}

The experimental transmission spectra are shown in Fig.~\ref{expbsc}, demonstrating
dramatic tuning of the resonance frequency. Furthermore, comparison of the experimental resonance
shift with the theoretical predictions shows (Fig.~\ref{compar}) that the experimental system demonstrates
even higher efficiency. This effect can be explained by accounting for the mutual capacitance between
resonators, neglected in our theoretical calculations.
Indeed, for the broadside-like configuration of rings, mutual capacitance
between them is distributed along the whole circumference~\cite{Ricardo}. Clearly, when
the resonators are laterally displaced, the mutual
capacitance decreases, so that this effect is added up
to the increase of resonance frequency imposed by decreased inductive
coupling.

The examples analyzed above illustrate the practical feasibility of the proposed tuning concept.
Particular details and tuning patterns may differ depending on the specific structural elements
used to create metamaterials, however it is clear that the remarkable resonance shift can be realized
over a wide range of alternative geometries, including numerous resonator varieties and even fishnet
structures which are more popular for higher frequencies. Remarkably, the proposed tuning mechanism
is not specific for the microwave range used in the above examples: this can be scaled in size and frequency
as long as metamaterial description in terms of effective medium is applicable. And on the practical side,
tremendous efficiency of the structural tuning can be used in a host of applications such as sensors, filters,
switches and all kinds of devices where prompt and sensitive response to changing conditions is required.

In conclusion, we have proposed and confirmed experimentally a novel concept
for tunability of metamaterials through a continuous adjustment of the lattice structure.

This work was supported by the Australian Research Council.
M.L. acknowledges hospitality of Nonlinear Physics Center
and a support of the Spanish Junta de Andalusia (P06-TIC-01368).
M.G. acknowledges support from the Russian Academy of Sciences (OFN
Programm ``Physics of new materials and structures'').


\end{document}